\documentclass[journal]{vgtc}                          

\graphicspath{{figures/}{pictures/}{images/}{./}} 

\usepackage{times}                     

\usepackage{tabu}                      
\usepackage{booktabs}                  
\usepackage{lipsum}                    
\usepackage{mwe}                       

\usepackage{mathptmx}                  

\usepackage{siunitx}
\usepackage{enumitem}
\usepackage{tabularx}

\usepackage{graphicx}
\usepackage{subcaption}

\usepackage[acronym,shortcuts]{glossaries}
\newacronym{HCI}{HCI}{human-computer interaction}
\newacronym{HMD}{HMD}{head-mounted display}
\newacronym{SBSOD}{SBSOD}{Santa Barbara Sense of Direction Scale}
\newacronym{MOT}{MOT}{Multiple Object Tracking}
\newacronym{AR}{AR}{augmented reality}
\newacronym{VR}{VR}{virtual reality}
\newacronym{XR}{XR}{extended reality}
\newacronym{FOV}{FOV}{field of view}

\usepackage[svgnames]{xcolor}

\onlineid{0}

\vgtccategory{Research}

\title{Actionable Guidance Outperforms Map and Compass Cues in Demanding Immersive VR Wayfinding}

\author{%
  \authororcid{Apurv Varshney}{0009-0000-7732-1249},
  \authororcid{Lily M. Turkstra}{0009-0009-9669-9189},
  Jiaxin Su,
  \authororcid{Mable Zhou}{0009-0000-7754-8574},\\
  \authororcid{Scott T. Grafton}{0000-0003-4015-3151},
  \authororcid{Barry Giesbrecht}{0000-0002-1976-1251},
  \authororcid{Mary Hegarty}{0000-0002-7777-8844}, and
  \authororcid{Michael Beyeler}{0000-0001-5233-844X}%
}

\authorfooter{
\item
All authors are with the University of California, Santa Barbara.
\item
Apurv Varshney and Lily M.\ Turkstra contributed equally to this work.
\item
Michael Beyeler is the corresponding author. E-mail: mbeyeler@ucsb.edu.
}

\teaser{
  \centering
  \includegraphics[width=0.8\linewidth]{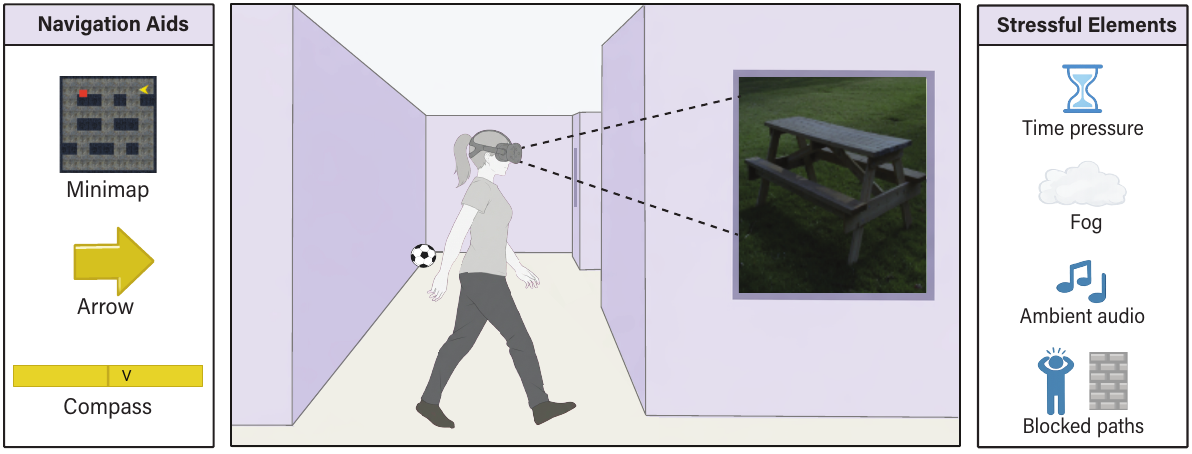}
  \caption{Overview of the immersive VR navigation task.
    \emph{Left:} Navigation aid conditions presented during the task: minimap, arrow, and compass. 
    \emph{Middle:} Schematic of in-world view during maze navigation. Participants physically walked through the environment to locate target landmarks. Dashed lines indicate the participant's visual field, and the soccer ball denotes an incidental attention probe used to measure selective attention during navigation.
    \emph{Right:} Environmental elements used to increase task demands, including time pressure, fog, ambient audio, and dynamic obstacles. In every trial, a wall descended as participants entered the first corridor, blocking the initially available route and forcing online replanning.
    }
  \label{fig:teaser}
}

\abstract{
Navigation aids are central to immersive virtual reality (VR) experiences that involve physical locomotion. Their effectiveness depends not only on how much spatial information they provide, but also on how directly that information supports movement decisions. We compared three common guidance techniques for immersive VR wayfinding: a directional arrow, a minimap, and a compass. In a controlled room-scale VR study with 42 participants completing 1008 trials, participants navigated to target landmarks in a time-pressured maze with reduced visibility and forced route replanning. 
Across behavioral and eye-tracking measures, arrow guidance produced the strongest navigation performance, minimap guidance yielded intermediate performance, and compass cues performed worst, suggesting that during immersive locomotion users benefit from guidance that can be interpreted rapidly while moving. 
These results suggest that in demanding immersive locomotion tasks, interfaces that translate spatial information directly into actionable movement cues can outperform richer but more interpretive spatial representations. Our findings highlight the importance of designing XR navigation interfaces that minimize the cognitive translation between spatial information and movement decisions.
} 


\begin{document}


\firstsection{Introduction}

\maketitle

Wayfinding is a core interaction in immersive \ac{XR}~\cite{bowman_travel_1997,LaViola2017_3DUI2e}. Navigation aids differ not only in how much spatial information they provide, but also in how much effort users must expend to translate that information into action~\cite{DarkenCevik1999_IEEEVR_MapOrientation,Kotlarek2018_SUI_SpatialOrientation,Chittaro2004_AVI_3DArrows,Kraus2020_ISMAR_OrientationTools}. This distinction becomes especially important during room-scale locomotion, where users must make navigation decisions while walking, monitoring obstacles, and maintaining spatial orientation under continuous perceptual load~\cite{Lee2022_IEEEVR_MRNavPreference,Kumaran2023_CHI_ARNavAids,marsh_cognitive_2013,sheik-nainar_influence_2015}. Under such conditions, richer spatial representations do not necessarily translate into better guidance~\cite{abdurrahman_cognitive_2022,gardony_how_2013}.

This distinction matters because common navigation aids place different interpretive demands on the user~\cite{Kraus2020_ISMAR_OrientationTools}. 
Directional arrows provide immediate egocentric guidance toward a goal~\cite{Chittaro2004_AVI_3DArrows}. 
Minimaps provide a spatial overview of the environment, requiring users to interpret the map and align it with their viewpoint~\cite{DarkenCevik1999_IEEEVR_MapOrientation,Kotlarek2018_SUI_SpatialOrientation, bowman_maintaining_1999}. 
Compasses provide only goal direction, leaving users to determine a feasible route~\cite{Kraus2020_ISMAR_OrientationTools,Lee2022_IEEEVR_MRNavPreference,Kumaran2023_CHI_ARNavAids}. 
Consequently, these interfaces differ not only in the information they provide but also in the cognitive demands they impose during navigation.

These demands become especially consequential during immersive locomotion. 
Prior work shows that navigation interfaces can shape not only wayfinding success but also visual attention and situational awareness~\cite{abdurrahman_cognitive_2022,bowman_maintaining_1999, quinn_augmented_2024}. 
For example, navigation aids in mobile and wide-area \ac{AR} can affect users' ability to locate targets and recall surrounding objects~\cite{Kumaran2023_CHI_ARNavAids, kumaran_scene_2025}. 
Other work shows that the placement and anchoring of guidance cues influence how users allocate visual attention during navigation~\cite{Peereboom2024_VR_AnchoringDR,Wang2025_TVCG_BodyFixedGuidance}. 
Navigation behavior can also shift under demanding conditions such as time pressure or uncertainty, increasing reliance on simpler and more reactive forms of guidance~\cite{brunye_risk-taking_2019,varshney_stress_2024, yavuz_determinants_2026,huang_effect_2026, kelley_guiding_2024, kelley_importance_2025,zhou_adaptive_2026}. 
Together, these findings suggest that the effectiveness of a navigation aid depends not only on the information it provides but also on how readily that information supports action during locomotion.

Despite their widespread use, many immersive navigation interfaces are still guided more by convention than by direct empirical comparison. 
Prior work has examined arrows~\cite{Chittaro2004_AVI_3DArrows}, minimaps~\cite{DarkenCevik1999_IEEEVR_MapOrientation,Kotlarek2018_SUI_SpatialOrientation}, compass-style tools~\cite{Kraus2020_ISMAR_OrientationTools,Lee2022_IEEEVR_MRNavPreference}, and related aids in mobile \ac{AR}~\cite{Kumaran2023_CHI_ARNavAids, mulloni_user_2011}. 
However, these techniques are often evaluated in isolation or in different tasks and environments. 
As a result, it remains unclear how actionable guidance compares with richer but more interpretive spatial aids within the same immersive wayfinding scenario.

In this paper, we directly compare three widely used guidance techniques in a controlled room-scale \ac{VR} navigation task: a directional arrow, a minimap, and a compass (\cref{fig:teaser}).
The arrow provides a direct action cue by indicating the egocentric direction of the goal. Unlike world-anchored arrows that guide users along a path~\cite{Kumaran2023_CHI_ARNavAids}, our view-fixed arrow indicates target direction even when no feasible route is directly visible, allowing users to resolve local route choices while progressing toward the goal.
In contrast, the compass provides only goal bearing, requiring users to align their heading (``$|$'' in \cref{fig:teaser}, \emph{Left}) with the target direction (``V'') and determine a valid route through the maze.
The minimap provides an allocentric overview of the maze layout, requiring users to interpret the map and translate it into movement decisions during navigation.

In a study with 42 participants and 1008 total trials, participants navigated to target landmarks in a time-pressured maze with reduced visibility, forced route replanning, and environmental stressors. 
Our primary outcome measure was navigation performance, defined as a composite metric combining path efficiency and completion time. 
As secondary measures, we examined selective attention to incidental objects and perceived workload to assess potential tradeoffs between navigation efficiency and situational awareness or task demand. 
We additionally conducted exploratory analyses of gaze behavior and self-reported navigation strategies.

Across measures, we observed a consistent ordering: arrow guidance produced the strongest navigation performance, minimap guidance provided intermediate support, and compass cues performed worst. 
This pattern suggests that in demanding immersive locomotion tasks, reducing the interpretive steps between interface information and movement decisions may improve navigation efficiency.

Our contribution is a controlled, within-task evaluation of three widely used navigation aids under the same physically embodied and time-pressured \ac{VR} wayfinding task. Specifically:
\begin{itemize}[topsep=0pt,itemsep=0pt,parsep=1pt]
    \item We present a controlled comparison of three common navigation aids (arrow, minimap, and compass) in a room-scale \ac{VR} wayfinding task involving time pressure, reduced visibility, and forced route replanning.
    \item We show that directional arrow guidance improves navigation performance relative to minimap and compass cues in this demanding immersive navigation scenario, while requiring less interface inspection and lower subjective workload than compass-based guidance.
\end{itemize}
By combining navigation performance with gaze, workload, stress, and incidental-object measures, the study characterizes not only which aid performs best, but also how each interface shapes attention and task demand during locomotion.

\section{Related Work}

\subsection{Navigation and Attention-Guiding Interfaces in XR}

A wide range of navigation aids have been proposed to support wayfinding in virtual and augmented environments~\cite{quinn_augmented_2024}. Early work in \ac{VR} examined map-based navigation interfaces and showed that factors such as map orientation relative to the user can strongly influence navigation performance~\cite{DarkenCevik1999_IEEEVR_MapOrientation}. Minimaps remain a common technique for providing spatial overview and supporting route planning in virtual environments~\cite{Kotlarek2018_SUI_SpatialOrientation,bowman_maintaining_1999,quinn_augmented_2024}.

Other approaches provide lightweight directional guidance rather than full spatial representations. 
For example, directional arrows embedded in the environment can guide users toward targets without explicitly representing the surrounding layout~\cite{Chittaro2004_AVI_3DArrows, kelley_guiding_2024}. 
In many augmented and mixed reality systems, such cues are anchored in the environment as path indicators or spatial overlays that guide users along a route~\cite{Kumaran2023_CHI_ARNavAids}. 
Compass-style indicators instead provide only the bearing of the goal relative to the user, leaving route planning to the user~\cite{Kraus2020_ISMAR_OrientationTools,Lee2022_IEEEVR_MRNavPreference}. 
In contrast to the world-anchored arrows used in much of this prior work, the arrow used in our study is a view-fixed interface element that continuously indicates the egocentric direction of the goal rather than prescribing a path through the environment.

While these approaches demonstrate the usefulness of various guidance techniques, they are often evaluated independently or in different task contexts. 
As a result, it remains difficult to determine how commonly used navigation aids compare when users must navigate the same immersive environment under identical conditions.

\subsection{Attention and Cognitive Demands During Immersive Navigation}

Navigation in immersive environments requires users to allocate attention across multiple competing demands, including movement control, obstacle monitoring, spatial orientation, and task goals~\cite{varshney_stress_2024, bowman_maintaining_1999, sheik-nainar_influence_2015, kumaran_scene_2025, kim_investigating_2022}. 
Interfaces that present additional visual information can therefore influence not only navigation success but also how users distribute attention during locomotion~\cite{gardony_how_2013, abdurrahman_cognitive_2022}. 
Prior work has shown that the placement and embodiment of visual cues in head-worn displays can strongly influence glance behavior and spatial knowledge acquisition~\cite{de2020place}, along with general wayfinding behavior and spatial memory~\cite{ishikawa_relationships_2013}.

Additional visual information presented via navigation aid cues has been shown to garner similar effects, such as shaping visual attention and situational awareness during immersive navigation~\cite{gardony_how_2013, quinn_augmented_2024}. 
For example, navigation cues in mobile and wide-area \ac{AR} can affect users' ability to detect and remember objects in the surrounding environment~\cite{Kumaran2023_CHI_ARNavAids, kim_investigating_2022}. 
Other work demonstrates that the placement and anchoring of guidance cues influences gaze behavior and visual attention during navigation tasks~\cite{Peereboom2024_VR_AnchoringDR,Wang2025_TVCG_BodyFixedGuidance}. 
More broadly, augmentations presented while users are moving can compete for perceptual resources and reduce awareness of the surrounding environment~\cite{kim_go_2025}.
Although navigation aids can improve wayfinding performance, they can also divide attention, impair spatial learning, and reduce reliance on internal navigation strategies~\cite{gardony_how_2013,ishikawa_maps_2016,bowman_maintaining_1999,ruginski_gps_2019}.
Designers must therefore balance potential performance benefits of navigation aids against possible negative impacts such as divided attention, reduced spatial learning, or increased cognitive load.~\cite{setu_predicting_2025}. 
Related immersive-VR work similarly shows that how task-relevant information is presented can affect wayfinding success, collision avoidance, and perceived difficulty, highlighting tradeoffs among information density, task relevance, and perceptual clarity~\cite{kasowski_static_2025}.

Navigation behavior can also shift under demanding conditions such as time pressure, uncertainty, or environmental stressors~\cite{yavuz_determinants_2026,varshney_stress_2024,huang_effect_2026, kelley_guiding_2024, kelley_importance_2025}. 
Under these circumstances, users may rely more heavily on simple and reactive navigation strategies, or previously learned routes rather than constructing or consulting detailed spatial representations~\cite{brunye_risk-taking_2019,varshney_stress_2024}. 
These findings suggest that the effectiveness of a navigation aid may depend not only on the information it provides, but also on the cognitive effort required to translate that information into movement decisions during locomotion~\cite{brugger_how_2019}.
This distinction reflects a broader design tradeoff between representational richness and actionability in navigation interfaces.

\subsection{Research Gap}

Despite extensive research on individual navigation aids, direct comparisons between commonly used guidance techniques remain limited in immersive locomotion settings. Arrows, minimaps, and compass-style indicators have largely been studied in isolation or under different tasks, leaving it unclear how these interfaces compare when users navigate the same immersive environment while walking and making decisions under perceptual load.

In particular, little empirical work has examined how aids that provide immediate directional guidance compare with aids that require users to interpret spatial representations during locomotion. This distinction is especially important in immersive \ac{XR}, where users must continuously translate interface information into movement decisions.

To address this gap, we conduct a controlled immersive \ac{VR} study directly comparing three widely used navigation aids: a directional arrow, a minimap, and a compass.

\section{Methods}

Participants first completed individual-difference measures and VR familiarization, then learned the maze along a fixed guided route. They subsequently completed an Unaided block followed by counterbalanced Arrow, Minimap, and Compass blocks, each containing one practice trial and six test trials. Across test trials, start and goal locations varied, visibility was limited by fog, and a blocked corridor forced online replanning. We measured navigation performance, incidental-object detection, perceived stress and workload, gaze behavior, and self-reported navigation strategies.

\subsection{Participants}

Forty-two participants (mean age = 19.6 years; 67\% female) were recruited from the university community and received course credit or a \$10/hour gift card. The study was approved by the Institutional Review Board (IRB) of the University of California, Santa Barbara (Protocol No.\ 68-26-0182), and all participants provided informed consent.

Approximately 60\% of participants reported limited prior VR experience (1–5 prior sessions), and 25\% reported no prior VR exposure. All participants had normal or corrected-to-normal vision.

Motion sickness was low (M = 1.37, SD = 0.98 on a 7-point scale), and no participants withdrew.

\begin{figure}[!t]
    \centering
    \includegraphics[width=\linewidth]{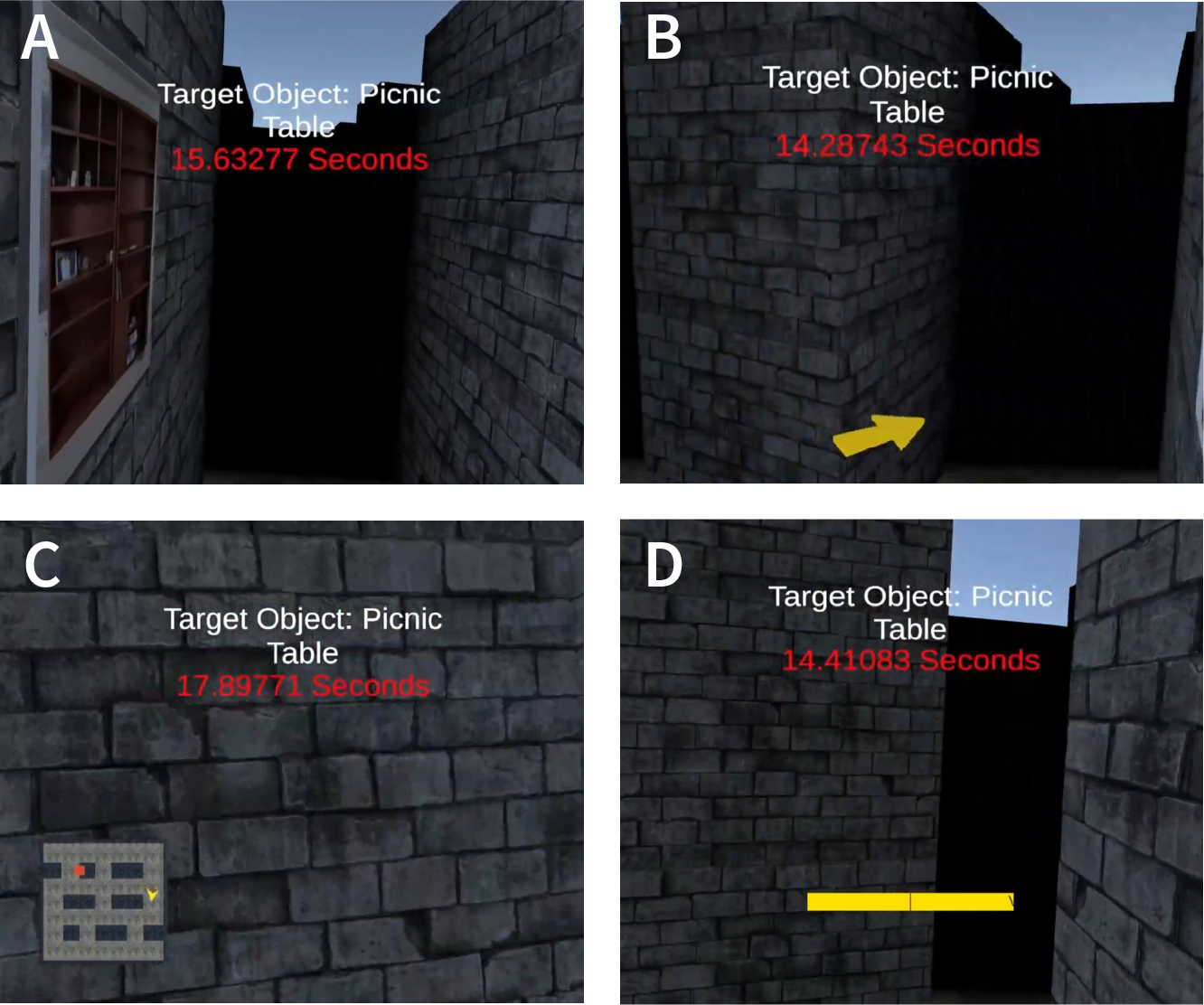}
    \caption{Representative first-person views during navigation trials.
    (A) Unaided navigation. 
    (B) Arrow guidance providing an egocentric directional cue toward the goal (which may point through walls when the target is occluded). 
    (C) Minimap showing a north-up overview of the maze with participant and target locations. 
    (D) Compass indicating the goal bearing relative to the participant's current heading. 
    Images show the right-eye view captured from the SteamVR compositor during navigation trials.}
    \label{fig:methods-hmd-view}
\end{figure}

\subsection{Equipment}

The immersive environment was presented using an HTC Vive Pro Eye wireless head-mounted display (1140 $\times$ 1600 pixels per eye, \SI{90}{\hertz}, $\sim$\SI{110}{\degree} field of view). Headset position and orientation were tracked using Lighthouse base stations within a $\SI{7.5}{\meter} \times \SI{7.5}{\meter}$ room-scale space.
The experiment was implemented in Unity and ran on a desktop computer with an Intel i7 CPU and NVIDIA RTX 3080 Ti GPU. 
Participants held a Vive controller in their dominant hand to confirm target selections and respond to in-VR questionnaires.
The headset's integrated Tobii eye tracker (\SI{120}{\hertz}) recorded gaze during navigation. Participants completed a standard eye calibration procedure prior to the experiment.
Spatial audio was delivered through the headset's integrated headphones.

\begin{figure}[!t]
    \centering
    \includegraphics[width=\linewidth]{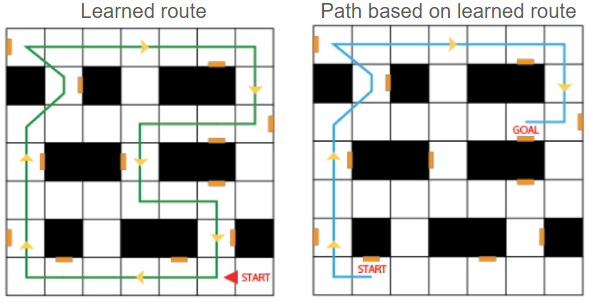}
    \caption{Top-down schematic of the VR maze. Landmarks (paintings) are denoted by orange nodes; walls are shown as black tiles. (A) The green path represents the guided learning route. (B) The blue path represents an example path during a navigation trial, while yellow triangles represent path directionality. }
    \label{fig:methods-maze}
\end{figure}

\subsection{Navigation Aids}

We compared three navigation aids representing common guidance strategies in immersive \ac{XR}: a directional arrow, a minimap, and a compass (\cref{fig:teaser}; see Supplemental Video). These aids differ in how directly interface information translates into movement decisions: the arrow provides an explicit directional cue toward the goal (a specified target painting in the virtual maze), the compass provides goal bearing only, and the minimap provides an allocentric overview of the environment.

All aids were continuously visible during navigation trials (\cref{fig:methods-hmd-view}). The arrow and compass appeared near the lower center of the field of view, whereas the minimap was placed in the lower-left corner to reduce occlusion of the scene.

\paragraph{Arrow}
The arrow aid provided a continuous egocentric cue toward the goal. A view-fixed arrow appeared near the lower center of the field of view and rotated to indicate the direction of the target~\cite{Chittaro2004_AVI_3DArrows,LaViola2017_3DUI2e}. Because the arrow encoded only goal direction rather than a feasible path, it could point through walls when the target was occluded.

\paragraph{Minimap}
The minimap displayed a north-up overview of the maze layout in the lower-left field of view, including the participant’s position and the target location. Minimap-style navigation aids are widely used in virtual environments and games to provide allocentric spatial context~\cite{DarkenCevik1999_IEEEVR_MapOrientation,Kotlarek2018_SUI_SpatialOrientation}. Because the map was head-fixed, participants had to translate allocentric information into egocentric movement decisions.

\paragraph{Compass}
The compass provided goal bearing without revealing the maze layout. A horizontal compass bar at the bottom of the field of view indicated the target heading relative to the participant’s current orientation, a common orientation cue in XR navigation systems~\cite{Kraus2020_ISMAR_OrientationTools,Lee2022_IEEEVR_MRNavPreference}. Participants therefore had to align their heading with the indicated direction while determining a feasible route through the maze.

\subsection{Environment and Task}
\label{sec:methods-task}

Participants navigated a room-scale virtual maze while physically walking within the tracked environment  (\cref{fig:methods-maze}).
The maze consisted of corridors and intersections connecting twelve target landmark locations marked by unique paintings, all of which remained present throughout every trial.
At the start of each trial, participants spawned at a randomized location and navigated to a specified target landmark before the time limit expired.
Start and goal locations varied across trials, and the six test trials within each condition used nonrepeating start--goal pairs.

\paragraph{Environmental Stressors}
Each trial included several constraints to increase navigation difficulty. Participants had \SI{20}{\second} to reach the target painting while a visible countdown timer remained on screen. Dense fog limited visibility to approximately \SI{2.5}{\meter}, so participants navigated within a stable landmark environment but had limited access to long-range visual cues.

At the first intersection after the start location, entering one of the available corridors triggered a wall that descended with a loud crashing sound, blocking that route and forcing replanning. 
Each trial contained exactly one such blocked corridor.
Unsettling ambient audio played throughout the trial to increase task tension.

\paragraph{Incidental Objects}
Twelve salient everyday objects were placed at fixed locations along potential routes through the maze. These objects were irrelevant to the navigation goal and served as a selective-attention probe: the navigation measures quantified efficiency, whereas the object task tested whether aid-related performance benefits were accompanied by reduced awareness of task-irrelevant environmental content. This tradeoff is relevant in XR because users following guidance may still need to notice people, obstacles, hazards, or contextual cues. Object identities were randomized across locations on each trial. Afterward, each participant's traversed route was used to determine which objects had been encountered, allowing detection sensitivity to distinguish encountered from non-encountered objects.

\begin{figure}[!t]
    \centering
    \includegraphics[width=0.9\linewidth]{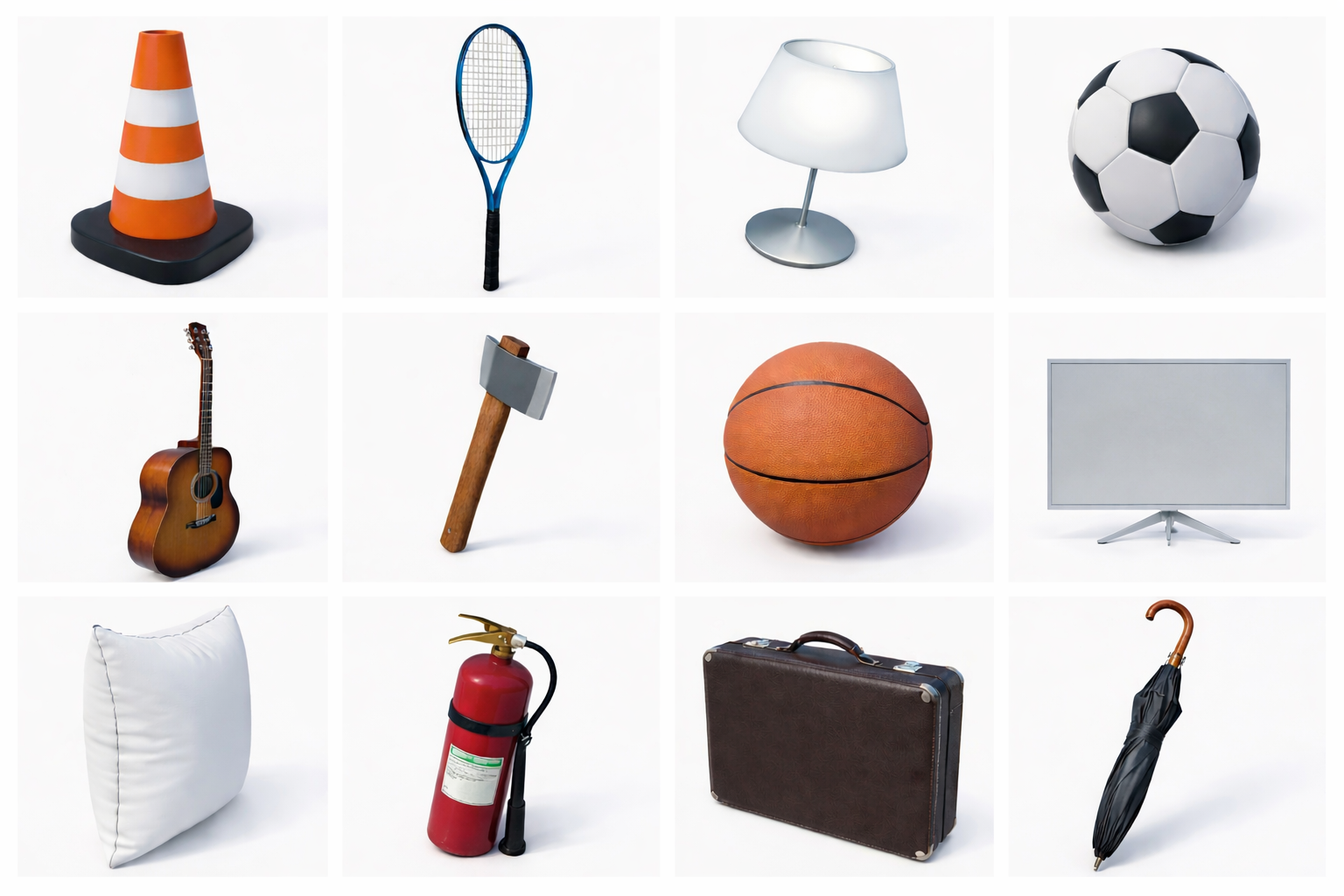}
    \caption{Incidental objects used for the secondary selective attention task.
    Twelve visually salient everyday objects were placed along potential paths in the maze during navigation trials. 
    These objects were not relevant to the navigation task but served as incidental stimuli to measure participants’ attention to the surrounding environment. 
    After each trial, participants reported which objects they noticed while navigating.}
    \label{fig:methods-2nd-objects}
\end{figure}

\subsection{Procedure}

The experiment consisted of three phases: pre-experiment measures, VR familiarization and environment learning, and the navigation experiment.

\paragraph{Pre-Experiment Measures}
First, participants completed a multiple-object tracking task to assess attentional tracking ability. Participants were then fitted with the VR headset and the eye tracker was calibrated.

\paragraph{VR Familiarization and Environment Learning}
Participants first explored a separate, neutral VR maze to become familiar with the headset, controller, room-scale walking, and item selection before beginning the navigation experiment.

They were then introduced to the experimental maze through a guided learning phase. Sequential, static red arrow markers indicated a fixed route through the environment. These markers were visually and functionally distinct from the experimental Arrow aid, which was yellow, dynamic, continuously visible, and indicated the direction of the current target rather than tracing a route. The learning markers were intended to establish basic maze familiarity and comfort with room-scale navigation, not to train participants on the experimental Arrow interface.

Participants repeated the guided route several times and then retraced it without guidance until they could complete it successfully. Experimental trials used varied start--goal pairs, fog, time pressure, and a blocked corridor that forced online replanning; they therefore could not be completed by simply following the learned route. The familiarization procedure and the distinction between the learning markers and experimental Arrow aid are also shown in the Supplemental Video.

\paragraph{Navigation Experiment}
Participants navigated under four conditions: Arrow, Minimap, Compass, and Unaided. Each condition included one practice trial followed by six test trials.
Because Unaided was always presented first to avoid aid-driven learning before the baseline measurement, including it in the inferential models would confound condition with practice and environment learning. We therefore treated Unaided as a descriptive baseline and restricted inferential comparisons to the three counterbalanced aid conditions. Trial order was included as a fixed effect in the trial-level models.

\begin{figure*}[!t]
    \centering
    \includegraphics[width=\linewidth]{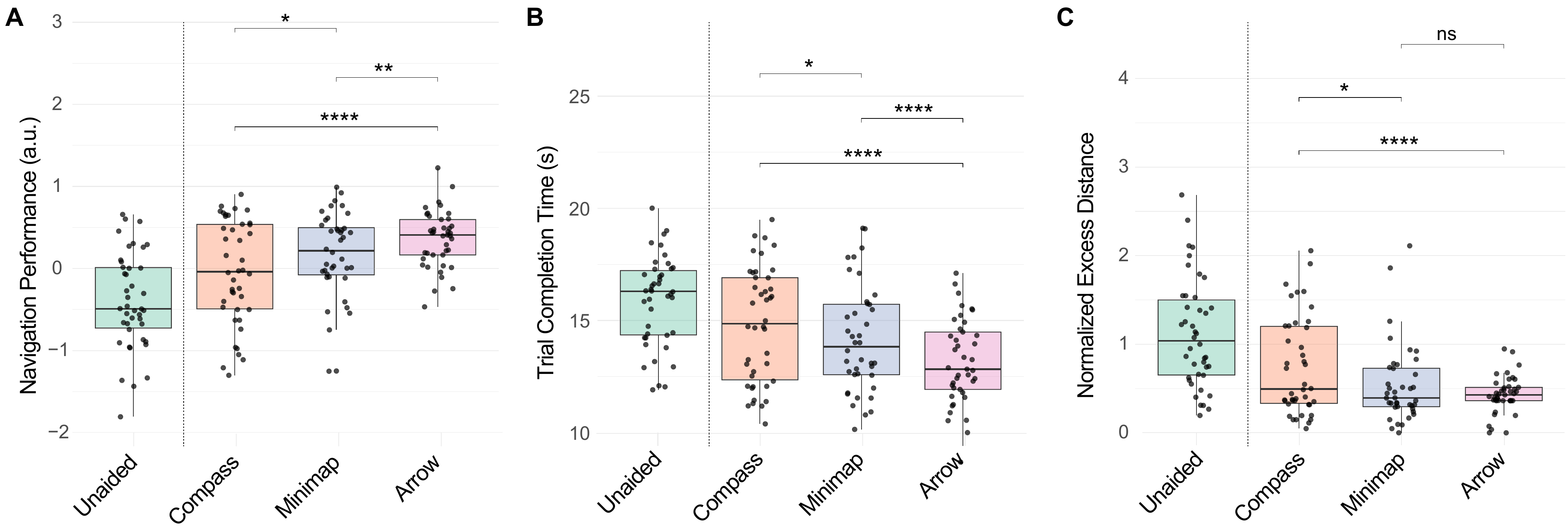}
    \caption{\textbf{Primary navigation performance across aids.}
    (A) Composite navigation performance score (\texttt{Nav\_Comp}), combining completion time and excess distance (higher values indicate better performance). Arrow outperformed both Minimap and Compass. 
    (B) Trial completion time (s). Arrow yielded faster completion time than Minimap and Compass. Times are shown on the original scale for interpretability; statistical analyses used log-transformed completion time. 
    (C) Normalized excess distance relative to the optimal path. All selected start--goal pairs had a shortest-path distance of at least \SI{9}{\meter}. While Arrow and Minimap reduced excess distance relative to Compass, their difference was not significant. 
    The vertical dashed line separates the Unaided baseline from the navigation-aid conditions; the Unaided block was always presented first and is shown for descriptive comparison only. 
    Points represent individual participants and boxplots summarize condition distributions. 
    Asterisks denote significance levels (ns $p \ge .05$, * $p < .05$, ** $p < .01$, **** $p < .001$).}
    \label{fig:results-primary}
\end{figure*}

\subsection{Measures}
\label{sec:methods-measures}

We collected objective navigation performance metrics, secondary behavioral measures, subjective workload ratings, and qualitative reports of navigation strategy.

\paragraph{Navigation Performance}
Our primary dependent variable was a composite navigation performance score (\texttt{Nav\_Comp}) combining excess distance and completion time.

To compute excess distance relative to the optimal path, the maze was discretized into \SI{1}{\meter} $\times$ \SI{1}{\meter} tiles corresponding to the corridor layout. 
All selected start--goal pairs had a shortest-path distance of at least \SI{9}{\meter}. Traversable tiles formed nodes in a graph, and walls were treated as non-traversable cells. For each trial, the optimal path length ($S$) was computed using Breadth First Search (BFS). Participant path length ($W$) was measured as the number of nodes traversed.
Excess distance was then defined as $(W-S)/S$, where $0$ indicates an optimal route and larger values indicate less efficient navigation.

Completion time was measured from trial onset to target selection. Excess distance and completion time were standardized across trials and combined so that higher values of \texttt{Nav\_Comp} reflected better navigation performance.

\paragraph{Situational Awareness}
Awareness of objects in the environment during navigation was assessed using the incidental objects placed throughout the maze (\cref{fig:methods-2nd-objects}). 
After each trial, participants were shown the set of possible objects and indicated which ones they had noticed while navigating. Previous studies have shown that during AR search tasks that are more difficult, awareness of objects in the environment is reduced \cite{kim_investigating_2022, Kumaran2023_CHI_ARNavAids}.
Based on the objects that appeared along the route traversed by the participant, responses were classified as hits, misses, false alarms, and correct rejections. 
Detection sensitivity ($d'$) was then computed for each navigation aid condition, with higher values indicating better discrimination between encountered and non-encountered objects.

\paragraph{Workload}

Subjective workload was measured using the NASA Task Load Index (TLX) survey, completed at the end of each aid block~\cite{hart_development_1988}. A composite workload score was computed as the mean of four TLX subscales (mental demand, physical demand, temporal demand, and effort), where higher values indicate greater perceived workload. TLX-4 was used to represent task load without conflation from frustration and performance subscale scores~\cite{tubbs-cooley_nasa_2018}.

\paragraph{Self-Reported Stress}

To verify that the environmental manipulations produced a demanding task, participants rated their perceived stress level after each trial using an in-VR scale ranging from 1 (``no stress'') to 7 (``very high stress'').

\paragraph{Strategy Reports}

After completing each navigation-aid block, participants provided open-ended descriptions of the strategies they used to navigate the maze. These responses were later analyzed using inductive thematic analysis to identify common navigation strategies.

\paragraph{Individual Differences}
We assessed attentional tracking capacity using a computerized \ac{MOT} task~\cite{pylyshyn_tracking_1988,meyerhoff_individual_2020,hulleman_mathematics_2005}, self-reported navigational ability using the 15-item \ac{SBSOD}~\cite{hegarty_development_2002}, and spatial anxiety using the 13-item Spatial Anxiety Scale~\cite{lawton_gender_1994,varshney_stress_2024}. The \ac{MOT} task was completed before the VR experiment; the questionnaires were completed afterward.

The \ac{MOT} task comprised five blocks of ten trials and lasted approximately 10--15 minutes. On each trial, participants tracked a subset of moving targets among identical distractors and then identified the targets. Performance was summarized using the task's estimated tracking-capacity score. Full task parameters, scoring procedures, questionnaire items, and response scales are provided in Supplemental Methods.

\paragraph{Eye Movements}
Eye movements were analyzed to quantify visual attention to the navigation interfaces. Because locomotion can reduce the accuracy of integrated eye trackers in some head-worn displays~\cite{awasthi_eye_2024}, we applied trial-level screening for gaze-data quality. Gaze samples with invalid tracking or blinks were removed. Trials were excluded if they contained an average of two or fewer valid samples per second of trial duration. Fixations were identified using Tobii's default fixation-classification algorithm.

Areas of interest (AOIs) were defined as rectangular regions enclosing each navigation interface element, with a small margin to account for tracking noise. Dwell time was computed as the cumulative fixation duration within the corresponding AOI during each trial.

\subsection{Statistical Analysis}
\label{sec:methods-stats}

All analyses were conducted in R using linear mixed-effects models. The dataset contained 1008 navigation trials from 42 participants. For eye-tracking analyses, 53 trials were excluded due to brief tracking losses.

Participant was included as a random intercept in all models to account for repeated measurements. Trial-level outcomes (navigation performance and completion time) were analyzed at the trial level, whereas block-level measures (e.g., workload) were analyzed using one observation per participant per aid condition.

Because the unaided block was always presented first, it was treated as descriptive and excluded from inferential comparisons among navigation aids.

\paragraph{Navigation Performance}
Navigation performance (\texttt{Nav\_Comp}) was analyzed with a linear mixed-effects model including Aid and Trial Order as fixed effects and participant as a random intercept. Pairwise comparisons were computed using estimated marginal means with Tukey correction.

\paragraph{Completion Time}
Completion time was analyzed using a linear mixed-effects model with Aid and Trial Order as fixed effects. 
Because completion time exhibited a right-skewed distribution, values were log-transformed prior to analysis to improve residuals.

\paragraph{Selective Attention}
Detection sensitivity ($d'$) was analyzed at the block level using a mixed-effects model with Aid as a fixed effect.
Pairwise comparisons were evaluated using Tukey-adjusted estimated marginal means.

\paragraph{Workload}
NASA TLX-4 workload scores were analyzed at the block level using a mixed-effects model with Aid as a fixed effect.

\paragraph{Exploratory Analyses}
Several additional analyses were conducted to explore how participants interacted with the navigation interfaces. 
These included models relating visual dwell time on navigation aids to navigation performance and subjective workload, as well as analyses of individual-difference measures (spatial anxiety, \ac{SBSOD}, and \ac{MOT}). 
Trait variables were standardized prior to analysis. 
Degrees of freedom for fixed effects were estimated using Satterthwaite approximations.

Throughout the Results, model coefficients are distinguished from post-hoc pairwise contrasts. Pairwise comparisons are based on estimated marginal means with Tukey-adjusted $p$-values, and estimated marginal means are reported with 95\% confidence intervals.

\section{Results}

Analyses were organized into three tiers. Primary analyses examined navigation performance (\texttt{Nav\_Comp}), a composite of excess distance and completion time. Secondary analyses assessed potential tradeoffs with selective attention, stress, and workload. Exploratory analyses examined gaze behavior, navigation strategies, and individual differences.

\begin{figure*}[!t]
    \centering
    \includegraphics[width=\linewidth]{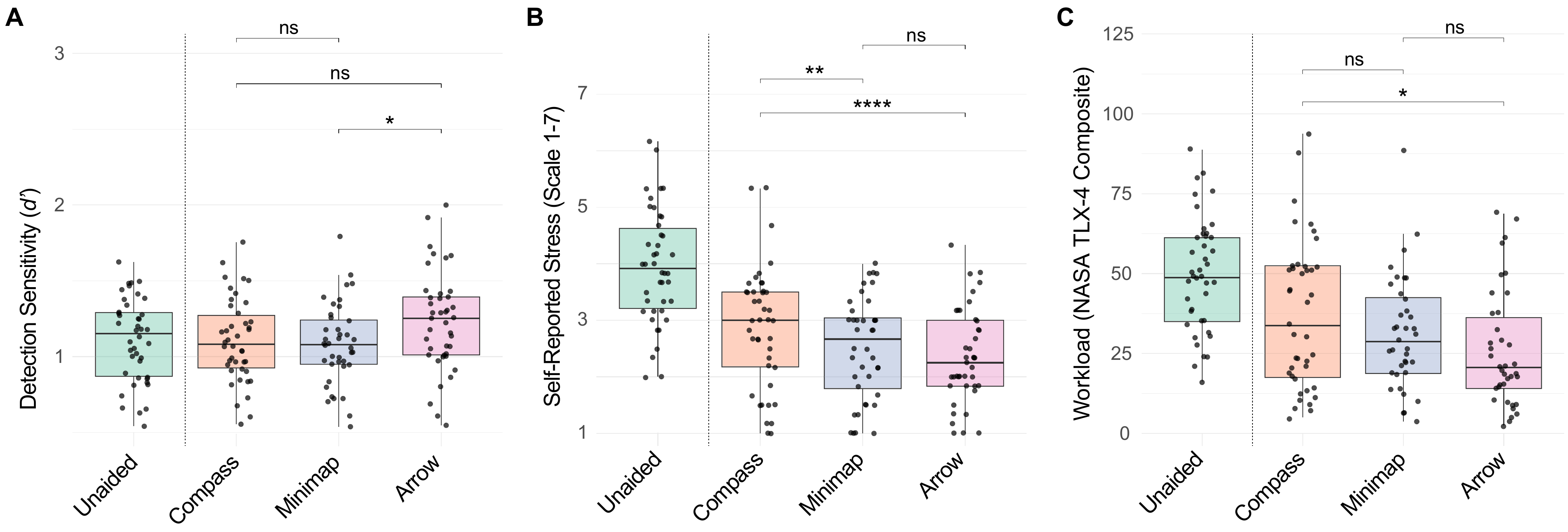}
    \caption{Secondary outcome measures across navigation aids.
    (A) Detection sensitivity ($d'$) for incidental objects. A modest difference was observed between Arrow and Minimap, with higher sensitivity under Arrow guidance, while other pairwise comparisons were not significant. 
    (B) Self-reported stress ratings during navigation trials. Compass produced higher stress ratings than both Arrow and Minimap. 
    (C) Subjective workload (NASA TLX-4). Arrow guidance produced lower workload than Compass, with other comparisons not reaching significance. 
    The vertical dashed line separates the Unaided baseline from the others; the Unaided block was always presented first and is shown for descriptive comparison only. 
    Points represent individual observations and boxplots summarize condition distributions.}
    \label{fig:results-secondary}
\end{figure*}

\subsection{Primary Outcome: Navigation Performance}

We first evaluated navigation performance using the composite metric \texttt{Nav\_Comp} (\cref{fig:results-primary}).

A linear mixed-effects model showed a significant effect of navigation aid on navigation performance. 
Relative to the Compass condition, both Minimap ($\beta = 0.162$, $SE = 0.053$, $t = 3.06$, $p = .002$) and Arrow ($\beta = 0.337$, $SE = 0.052$, $t = 6.51$, $p < .001$) significantly improved navigation performance. 
Pairwise comparisons with Tukey correction further showed that Arrow significantly outperformed Minimap ($\Delta = 0.175$, $SE = 0.054$, $t = 3.27$, $p = .003$).

Estimated marginal means reflected a clear ordering across aids (\cref{fig:results-primary}). 
Arrow produced the highest navigation performance (M = 0.49, 95\% CI [0.36, 0.61]), followed by Minimap (M = 0.31, 95\% CI [0.18, 0.44]), whereas Compass produced the lowest performance (M = 0.15, 95\% CI [0.02, 0.28]). 

We also observed a modest practice effect across trials ($\beta = 0.012$, $SE = 0.004$, $t = 3.00$, $p = .003$), indicating slight improvement with experience. 
The aid effects therefore remained after accounting for this modest improvement across trials.


\paragraph{Completion Time}
A linear mixed-effects model revealed an effect of navigation aid on log-transformed completion time (\cref{fig:results-primary}B). 
For visualization purposes, completion times are plotted on the original time scale in the figure.

Relative to Compass, participants completed trials faster when using Arrow ($\beta = -0.109$, $SE = 0.017$, $t = -6.33$, $p < .001$). 
The corresponding model coefficient for Minimap relative to Compass was smaller ($\beta = -0.037$, $SE = 0.018$, $t = -2.09$, unadjusted $p = .037$).
Trial order was also significant ($\beta = -0.0036$, $SE = 0.0014$, $t = -2.60$, $p = .0096$), again indicating modest improvement across trials.

Pairwise comparisons with Tukey correction confirmed that Arrow was significantly faster than both Compass ($\Delta = 0.109$, $SE = 0.017$, $t = 6.33$, $p < .001$) and Minimap ($\Delta = 0.072$, $SE = 0.018$, $t = 4.06$, $p < .001$). 
The difference between Compass and Minimap did not remain significant after correction ($\Delta = 0.037$, $SE = 0.018$, $t = 2.09$, $p = .093$).

Estimated marginal means showed the same ordering across conditions, with the fastest completion times under Arrow guidance ($\log M = 2.56$, 95\% CI $[2.51, 2.61]$), followed by Minimap ($\log M = 2.63$, 95\% CI $[2.58, 2.68]$), and Compass ($\log M = 2.66$, 95\% CI $[2.61, 2.71]$).

\paragraph{Excess Distance}

Excess distance analyses revealed a similar ordering across navigation aids (\cref{fig:results-primary}C). 
Participants traveled closest to the optimal path when using Arrow (Arrow: $M = 0.44$, 95\% CI $[0.31, 0.57]$; Minimap: $M = 0.55$, 95\% CI $[0.41, 0.68]$; Compass: $M = 0.74$, 95\% CI $[0.62, 0.87]$). 

A linear mixed-effects model revealed a significant effect of navigation aid on excess distance ($F(2, 667.98)=10.35$, $p < .001$). 
Pairwise comparisons with Tukey correction indicated that both Arrow ($\Delta = -0.303$, $SE = 0.067$, $t = -4.48$, $p < .001$) and Minimap ($\Delta = -0.197$, $SE = 0.069$, $t = -2.85$, $p = .012$) reduced excess distance relative to Compass, while the difference between Arrow and Minimap was not statistically significant ($p = .282$). 
%

\begin{figure*}[t!]
    \centering
    \includegraphics[width=\linewidth]{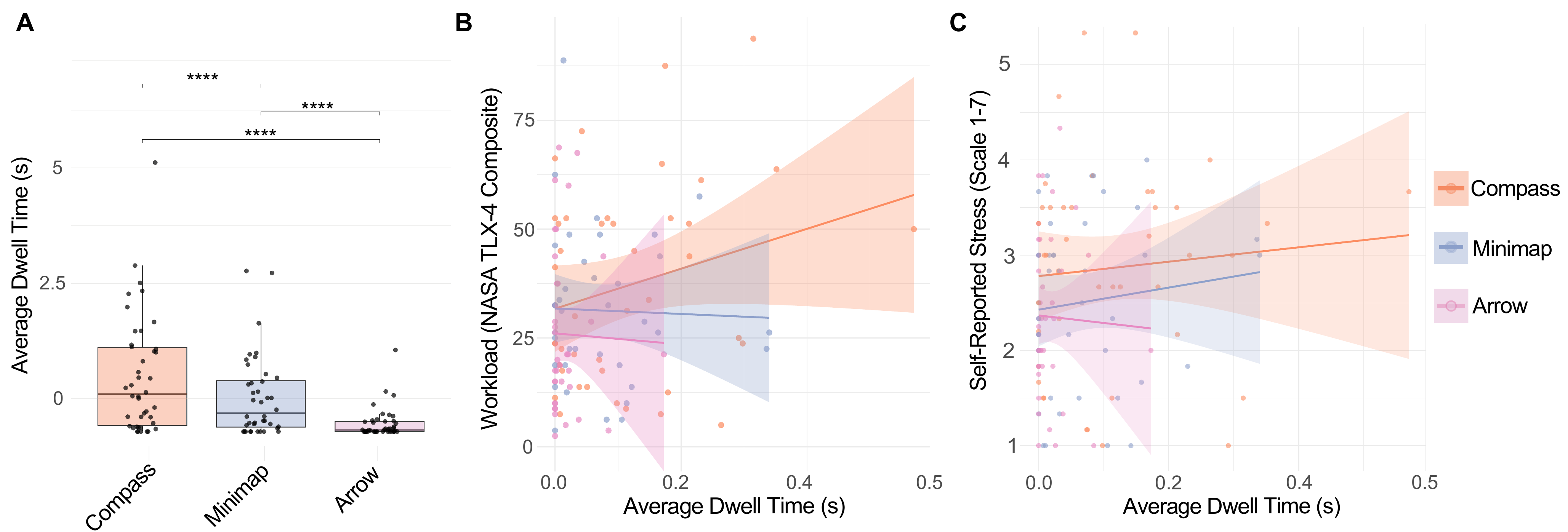}
    \caption{Impact of interface dwell time on workload and stress.
    (A) Average interface dwell time (s) per navigation aid. Arrow guidance required the least dwell time, followed by Minimap and Compass. 
    (B) Relationship between average dwell time (s) and subject-level composite NASA TLX-4 workload scores for each aid. 
    (C) Relationship between average dwell time (s) and per-trial self-reported stress ratings (1--7 scale).
    Points represent individual observations; colored lines indicate linear regression fits for each navigation aid condition; shaded regions denote 95\% confidence intervals.}
    \label{fig:results-dwell}
\end{figure*}

\begin{figure*}[!t]
    \centering
    \includegraphics[width=\linewidth]{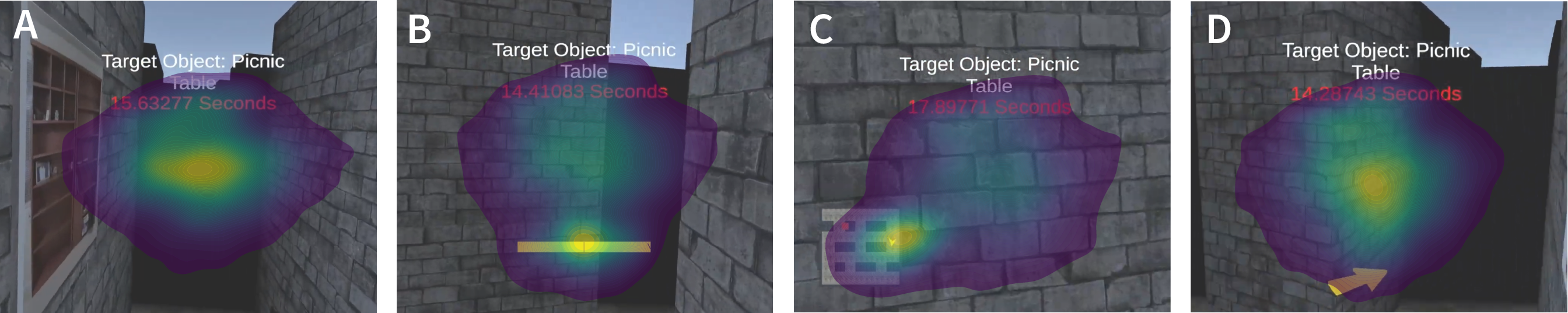}
    \caption{Aggregated gaze heatmaps across navigation aids.
    Kernel density estimates of fixation locations during navigation trials, aggregated across participants and trials (higher values indicate greater fixation density).
    Heatmaps are overlaid on the representative first-person views shown in \cref{fig:methods-hmd-view} to provide spatial reference for the navigation interface elements.
    (A) Unaided navigation, (B) Compass, (C) Minimap, and (D) Arrow guidance.
    Fixations concentrate near the interface elements for Compass and Minimap, whereas gaze is more broadly distributed across the scene under Arrow guidance.}
    \label{fig:results-gaze}
\end{figure*}

\subsection{Secondary Outcomes}

We next examined whether the improved navigation performance observed with arrow guidance came at the cost of reduced environmental awareness or increased cognitive demand (\cref{fig:results-secondary}).

\paragraph{Situational Awareness}
Improved navigation performance could potentially come at the cost of reduced attention to the surrounding environment. 
To evaluate this possibility, we examined detection sensitivity ($d'$) for encountered incidental objects. 
Because $d'$ depends on both hits and false alarms, it was computed at the block level for each participant and aid condition. A linear mixed-effects model revealed a significant effect of navigation aid on detection sensitivity ($F(2, 80.77) = 4.42$, $p = .015$).
Pairwise comparisons with Tukey correction showed higher sensitivity under Arrow than Minimap ($\Delta = 0.149$, $SE = 0.054$, $t = 2.76$, $p = .019$), while other pairwise differences were not significant after correction (Compass-Arrow, $p = .057$; Compass-Minimap, $p = .888$).

Estimated marginal means were highest in the Arrow condition ($M = 1.23$, 95\% CI $[1.14, 1.32]$), followed by Compass ($M = 1.11$, 95\% CI $[1.02, 1.19]$) and Minimap ($M = 1.08$, 95\% CI $[0.99, 1.17]$)(\cref{fig:results-secondary}A). 
Across conditions, detection sensitivity remained in the moderate range ($d' \approx 1.1$--$1.3$), indicating that participants were able to discriminate encountered from non-encountered objects well above chance while navigating the maze.

These results provide no evidence that the navigation benefit of Arrow guidance came at the expense of incidental-object awareness in this task.

\paragraph{Stress}
Participants also reported perceived stress after each trial (\cref{fig:results-secondary}B). 
Stress followed a pattern similar to workload. Estimated marginal means indicated the lowest stress in the Arrow condition ($M = 2.36$, 95\% CI $[2.09, 2.63]$), followed by Minimap ($M = 2.54$, 95\% CI $[2.27, 2.81]$), and Compass ($M = 2.85$, 95\% CI $[2.58, 3.12]$).
A linear mixed-effects model revealed a significant effect of navigation aid on perceived stress ($F(2, 663.25)=15.16$, $p < .001$).
Pairwise comparisons with Tukey correction showed that Arrow produced significantly lower stress than Compass ($\Delta = -0.492$, $SE = 0.090$, $t = -5.44$, $p < .001$), and Minimap also produced lower stress than Compass ($\Delta = -0.312$, $SE = 0.092$, $t = -3.37$, $p = .002$). The difference between Arrow and Minimap was not significant after correction ($\Delta = -0.180$, $p = .132$).


\paragraph{Workload}
Perceived workload was assessed using NASA TLX-4 scores (\cref{fig:results-secondary}C). 
Estimated marginal means indicated the lowest workload in the Arrow condition ($M = 25.9$, 95\% CI $[19.4, 32.3]$), followed by Minimap ($M = 31.3$, 95\% CI $[24.8, 37.8]$), and Compass ($M = 37.2$, 95\% CI $[31.0, 43.4]$).

Pairwise comparisons with Tukey correction showed that Arrow produced significantly lower workload than Compass ($\Delta = -11.34$, $SE = 4.49$, $t = -2.53$, $p = .036$). 
Differences between Minimap and Compass ($\Delta = -5.88$, $p = .400$) and between Minimap and Arrow ($\Delta = 5.46$, $p = .466$) were not statistically significant after correction.

These results suggest that guidance cues that more directly support movement decisions can reduce perceived task demand during immersive navigation.

\subsection{Exploratory Analyses}

To better understand how participants interacted with the navigation interfaces, we conducted exploratory analyses examining gaze behavior, self-reported navigation strategies, and individual differences. These analyses were not part of the primary hypothesis tests and are intended to provide insight into the mechanisms underlying navigation performance.

\begin{table*}[!t]
    \caption{%
    Themes identified in participants’ reported navigation strategies.
    Values indicate the number of responses coded with each theme for each navigation aid condition. Because multiple themes could be assigned to a single response, counts are not mutually exclusive and do not sum to the number of participants.%
    }
    \label{tab:strategies}
    \centering%
    \begin{tabularx}{\linewidth}{lXcccc}
        \toprule
        Theme & Description & Unaided & Minimap & Compass & Arrow \\
        \midrule
        Aid reliance & Participants relied primarily on the navigation aid & - & 21 & 6 & 24 \\
        Memory-based navigation & Participants navigated using remembered routes or landmarks & 39 & 8 & 15 & 3 \\
        Combined strategy & Participants used both memory and the aid & - & 6 & 11 & 6 \\
        Reduced cognitive load & Participants reported that the aid reduced effort & - & 8 & 8 & 7 \\
        \bottomrule
    \end{tabularx}%
\end{table*}

\paragraph{Interface Dwell Time}
\label{sec:results-dwell}

We examined how much visual attention each navigation aid required during navigation by measuring dwell time on the aid interface. Mean dwell time showed a clear descriptive ordering across conditions (\cref{fig:results-dwell}A): participants spent the most time inspecting the Compass interface, less time on the Minimap, and the least time on the Arrow aid.

To test whether interface inspection related to navigation outcomes, we examined the relationship between average dwell time and subjective workload (\cref{fig:results-dwell}B). Dwell time showed a small positive association with perceived workload ($r = .21$, $t(114)=2.30$, $p = .023$, 95\% CI $[.03, .38]$), indicating that, on average, participants who spent more time inspecting the interface reported slightly higher NASA TLX-4 scores.

A similar but weaker relationship was observed for perceived stress (\cref{fig:results-dwell}C). Dwell time was positively associated with stress ratings ($r = .12$, $t(702)=3.14$, $p = .002$, 95\% CI $[.04, .19]$), suggesting that longer interface inspection tended to occur during more demanding navigation trials.

Aggregated gaze heatmaps show the same pattern, with gaze concentrated on the Compass and Minimap interfaces but distributed more broadly across the environment when using the Arrow (\cref{fig:results-gaze}).

\paragraph{Navigational Strategies and Preferences}

After each block, participants reported their navigation strategies. Responses were coded via inductive thematic analysis into four themes describing how participants combined memory and navigation aids (\cref{tab:strategies}).

Clear differences in reported strategies emerged across navigation aids. With Arrow and Minimap, participants most often reported relying directly on the interface guidance. One participant using the Minimap reported that they "just followed the Minimap instead of trying to remember where the paintings were."

In contrast, participants navigating unaided or with the Compass more often reported memory-based strategies, such as recalling previously encountered landmarks or reconstructing routes from earlier exploration. For example, one participant in the unaided condition reported "[making] a relational mental map of the location of the paintings." 

Participants using the Compass frequently reported that the aid provided limited actionable guidance toward the goal. One participant stated that the Compass was ``the least helpful in navigating, I ignored it and used my initial memory to navigate through the maze.''

By comparison, Arrow users more often described reactive strategies, such as following the arrow direction and adjusting their paths when corridors became blocked, with one explaining: ``I relied more on the arrow than on my memory. When the wall stopped me, I could just use the arrow to go right around it.''

Overall, participants favored the arrow the most, with 29 (69\%) participants ranking it as their most preferred aid, whereas Compass and Unaided were most often ranked least preferred.
When reporting their least favorite condition, 20 (47.6\%) people selected control, 16 (38.1\%) chose compass, and 6 (14.3\%) people selected minimap.

These qualitative reports mirror the quantitative findings:  
Arrow was associated with greater reliance on the interface itself and was the most preferred, whereas more abstract directional cues, or lack thereof in the control, encouraged greater reliance on memory-based navigation and were ranked lower.

\paragraph{Individual Differences}
Finally, we examined whether navigation aids interacted with individual differences in spatial ability. Mixed-effects models tested interactions between navigation aid and spatial anxiety, sense of direction (\ac{SBSOD}), and multiple-object tracking capacity (\ac{MOT}). None of these traits significantly moderated the effectiveness of the navigation aids, suggesting that the relative advantage of Arrow over Minimap and Compass was consistent across participants.

\begin{figure}[!t]
    \centering
    \includegraphics[width=.9\linewidth]{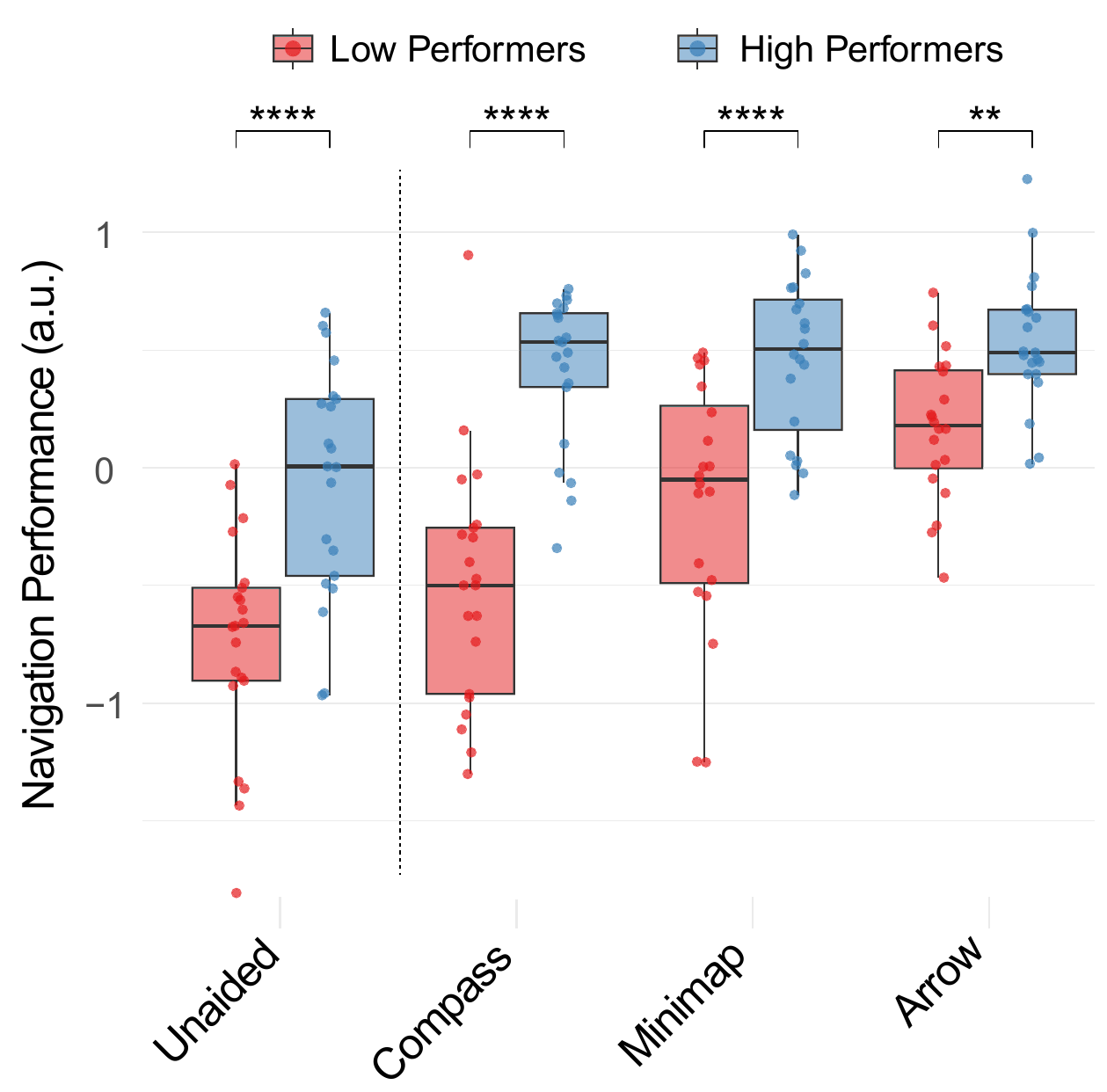}
    \caption{Navigation performance by performer group and navigation aid.
    Participants were divided into low and high performers using a median split of overall navigation performance. High performers achieved higher navigation performance across all conditions, but the magnitude of this difference varied across aids and was smallest in the Arrow condition. Points represent participant means; boxplots summarize group distributions.}
    \label{fig:results-ind-diff}
\end{figure}

To further explore potential differences between participants, we divided participants into high and low performers based on a median split of overall navigation performance. Descriptively, the performance gap between groups was largest in the Unaided condition and remained substantial for Compass and Minimap, whereas it was smaller under Arrow guidance (\cref{fig:results-ind-diff}). This exploratory pattern is consistent with the possibility that Arrow guidance may particularly benefit lower-performing participants, but it should be interpreted cautiously.

Post-hoc comparisons showed that the performance gap between high and low performers was largest in the Unaided condition and remained substantial for Compass and Minimap. In contrast, the difference between groups was smaller when participants used the Arrow aid (\cref{fig:results-ind-diff}). 
This exploratory subgroup pattern is consistent with the possibility that Arrow guidance may particularly benefit lower-performing participants, but it should be interpreted cautiously.

\paragraph{Gender}
We conducted exploratory analyses examining whether participant gender moderated the effect of navigation aid. We compared models with and without an Aid $\times$ gender interaction using likelihood-ratio tests. These tests indicated that gender moderated navigation performance ($\chi^2(2) = 8.37$, $p = .015$), completion time ($\chi^2(2) = 9.00$, $p = .011$), stress ($\chi^2(2) = 13.29$, $p = .001$), and dwell time ($\chi^2(2) = 17.90$, $p < .001$). In every case, the ordering of aids was preserved within both groups: Arrow yielded the best outcomes, followed by Minimap, then Compass. The interaction reflected differences in the estimated magnitude of the aid effects rather than a reversal of their ordering, with clearer separation among female participants and greater uncertainty among male participants.

These analyses were exploratory, and the sample was imbalanced by gender (27 female, 15 male). The wider uncertainty in the male subgroup is at least partly attributable to its smaller size. We therefore report these interactions descriptively and do not draw conclusions about gender-specific navigation behavior.


\section{Discussion}

This study examined how different forms of navigation guidance support wayfinding during immersive locomotion. By directly comparing three commonly used navigation aids (arrow, minimap, and compass) we evaluated how the representational form of guidance influences navigation behavior, visual attention, and perceived task demands. The results indicate that interfaces which translate spatial information into actionable movement cues can provide substantial advantages during demanding navigation tasks.

\subsection{Actionable Guidance Improves Navigation Performance}

The three navigation aids differed primarily in how directly they supported movement decisions. The Arrow aid provided a continuous directional cue indicating how to move toward the goal, whereas the Compass provided only orientation information and required users to infer a feasible route through the environment. The Minimap provided the most spatial information but required participants to translate an allocentric representation of the environment into egocentric movement decisions.

Our results indicate that the degree to which navigation cues translate spatial information into actionable movement has a strong impact on navigation performance. Participants using the Arrow aid consistently reached targets more efficiently and completed trials faster than those using the other interfaces. In contrast, the Compass aid often required participants to interpret directional information while simultaneously reasoning about the maze layout, which likely increased navigational uncertainty.

This pattern is broadly consistent with prior findings in augmented reality navigation systems, where users have been shown to prefer arrow-based guidance over other forms of navigation support~\cite{Kumaran2023_CHI_ARNavAids}. While that work focused primarily on user preference and subjective usability, our results extend these findings by demonstrating measurable improvements in navigation efficiency when guidance is presented as a direct directional cue.


\subsection{Visual Attention as a Window into Navigation Behavior}

Eye-tracking analyses provided additional insight into how participants interacted with the interfaces. Participants spent substantially less time inspecting the Arrow interface compared to the Minimap and Compass interfaces.
This reduced inspection suggests that the directional cue could be interpreted quickly during locomotion.

Moreover, longer dwell times on the interface were associated with higher subjective workload and stress. This pattern suggests that extended inspection of the navigation interface tended to occur during more demanding navigation episodes. In other words, participants appeared to consult the interface more frequently when additional guidance was needed to resolve route decisions.

Increased interface inspection was not associated with lower incidental-object detection. Thus, within the scope of this probe, greater reliance on the navigation interfaces was not accompanied by measurably poorer awareness of encountered objects. This measure captures one aspect of environmental awareness and should not be interpreted as a comprehensive test of situational awareness.

These results also highlight the value of eye movements as a behavioral window into how users interact with navigation interfaces. Prior work has explored gaze-driven adaptive aids for VR navigation~\cite{alghofaili_lost_2019}, and our findings suggest a particularly simple signal for such systems: prolonged interface inspection may indicate that users are struggling to interpret the aid or decide how to proceed.

\subsection{Implications for XR Navigation Interface Design}

These findings highlight the value of actionable guidance in immersive navigation interfaces. Interfaces that minimize the translation between spatial information and movement decisions may reduce cognitive overhead during locomotion and allow users to keep more attention on the surrounding environment~\cite{gardony_how_2013}.

Our gaze results are consistent with the concepts of information access cost and information access effort: consulting a display requires perceptual and cognitive effort, and users weigh that effort against the expected value of the information~\cite{poole2023information,poole2026information,warden2024information}. Participants spent more time looking at the Compass and Minimap than at the Arrow, suggesting that these interfaces required greater effort to extract and translate guidance into movement decisions. However, dwell time alone cannot distinguish interpretive difficulty from strategic reliance on the interface.

This interpretation is consistent with prior work showing that navigation tool design influences both immediate navigation behavior and how users later revisit virtual environments~\cite{yi_exploring_2023}. In immersive settings where users must make route decisions while moving and monitoring their surroundings, interfaces that provide simple, actionable cues may therefore offer substantial advantages.

The eye-tracking results further suggest that visual inspection of navigation aids may provide a behavioral signal of navigational difficulty. In future XR systems, such signals could potentially be used to adapt interface behavior dynamically, for example by modifying visual guidance when users appear to struggle with route decisions.

\subsection{Limitations and Future Work}

Several limitations should be considered. First, the controlled maze enabled systematic comparison of the aids but does not capture the scale, landmark richness, or planning demands of real-world navigation~\cite{setu2024mazed}. The fixed north-up minimap also required allocentric-to-egocentric transformation; heading-up maps may yield different results~\cite{DarkenCevik1999_IEEEVR_MapOrientation}.

Second, participants were drawn from a young university sample (mean age = 19.6 years), limiting generalizability. Sense of direction, spatial anxiety, and \ac{MOT} performance did not moderate the aid effects, but these exploratory measures do not replace broader sampling. Gender moderated several outcomes, although the aid ordering was unchanged and the sample was imbalanced (27 female, 15 male). These interactions should therefore be interpreted cautiously.

Third, Unaided was always presented first to preserve it as a baseline, confounding that condition with practice and environment learning; it was therefore treated descriptively. In addition, the guided learning phase used static red arrow markers. Although these differed from the yellow, dynamic experimental Arrow and traced a fixed route rather than indicating target direction, they may still have conferred some familiarity with arrow-like guidance. Future studies should use neutral or matched onboarding and a design that separates baseline measurement from environment learning.

Finally, future systems could use gaze or movement trajectories to detect navigational uncertainty and adapt guidance accordingly. Such approaches may help balance navigation efficiency with environmental awareness while minimizing distraction~\cite{davari_towards_2024,lu_evaluating_2021}.

\section{Conclusion}

We compared three common navigation aids in immersive environments: a directional arrow, a minimap, and a compass. Aids that translated spatial information into actionable movement cues produced the best navigation performance and required the least visual inspection. In contrast, aids requiring users to interpret directional or spatial information imposed greater attentional demands and yielded poorer outcomes. These results highlight the value of XR navigation interfaces that minimize the translation from spatial information to action.

\acknowledgments{
This research was sponsored by the U.S. Army Research Office and accomplished under contract W911NF19-D-0001 and cooperative agreement W911NF-19-2-0026 for the Institute for Collaborative Biotechnologies.
In accordance with IEEE guidelines, generative AI tools were used in limited ways during manuscript preparation. ChatGPT (OpenAI, GPT-5.3) was used for light figure editing (for \cref{fig:teaser} and \cref{fig:methods-2nd-objects}) and language refinement of portions of the manuscript. Gemini 3 Pro (Google) was used to assist with generating several data visualization scripts and figures in R. All AI-generated outputs were reviewed and verified by the authors, who take full responsibility for the content of the article.}


\section*{Supplemental Materials}
A Supplemental Video demonstrates the maze, guided learning phase, environmental stressors, and four navigation conditions. Supplemental Methods provide full details of the Multiple-Object Tracking task and the questionnaires.


\bibliographystyle{abbrv-doi-hyperref}

\bibliography{references}
\end{document}